\documentclass[a4paper,onecolumn,oneside,fleqn,10pt]{article}

\usepackage{abstract}
\usepackage{amsmath}
\usepackage{amssymb}
\usepackage{authblk}
\usepackage[english]{babel}
\usepackage{bm}
\usepackage{booktabs}
\usepackage[font={color=blue},figurename=Figure,labelfont={it}]{caption}
\usepackage{cite}
\usepackage{enumitem}
\usepackage{fchicago}
\usepackage{float}
\usepackage[left=2cm,right=2cm,top=2cm,bottom=2cm]{geometry}
\usepackage{graphicx}
\usepackage{hyperref}
\usepackage{indentfirst}
\usepackage[utf8]{inputenc}
\usepackage[switch,columnwise]{lineno}
\usepackage{microtype}
\usepackage{setspace}
\usepackage{subcaption}
\usepackage[T1]{fontenc}
\usepackage{newtxtext}
\usepackage{newtxmath}
\usepackage{wasysym}
\usepackage{fmtcount}
\usepackage{chngcntr}
\usepackage{titlesec}
\usepackage[table]{xcolor}
\usepackage{fancyhdr}
\usepackage{footmisc}
\usepackage{xspace}

\definecolor{blue}{RGB}{0,0,255}
\definecolor{matlabblue}{rgb}{0,0.4470,0.7410}

\hypersetup{
  colorlinks = true,
  linkcolor  = matlabblue,
  citecolor  = matlabblue,
  urlcolor   = matlabblue,
  filecolor  = matlabblue,
  linktoc    = all
}

\DeclareCaptionFont{bluefont}{\color{matlabblue}}
\newcommand{\eg}{\textit{eg. }}
\newcommand{\cf}{\textit{cf. }}
\newcommand{\ie}{\textit{ie. }}

\counterwithout{equation}{section}

\graphicspath{{figures/pdf/}{images/}{./}}

\newcommand{\includearticlefigure}[2][1\textwidth]{%
  \IfFileExists{figures/#2.pdf}{%
    \includegraphics[width=#1]{figures/#2.pdf}%
  }{%
    \IfFileExists{figures/#2.png}{%
      \includegraphics[width=#1]{figures/#2.png}%
    }{%
      \IfFileExists{#2.pdf}{%
        \includegraphics[width=#1]{#2.pdf}%
      }{%
        \IfFileExists{#2.png}{%
          \includegraphics[width=#1]{#2.png}%
        }{%
          \fbox{\parbox[c][6cm][c]{0.95\textwidth}{\centering Missing figure: #2}}%
        }%
      }%
    }%
  }%
}

\begin{document}

\title{\huge On the weak synchronization of a nonlinear solar dynamo oscillator by coherent planetary timing}

\author[1]{Le Mouël Jean-Louis\textsuperscript{\dag}}
\author[1]{Courtillot Vincent}
\author[2,3]{Kossobokov Vladimir}
\author[4]{Gibert Dominique}
\author[5]{Lopes Fernando}
\author[5]{Boulé Jean-Baptiste}
\author[6]{Zuddas Pierpaolo}

\affil[1]{\small Académie des Sciences, Institut de France, Paris, France}
\affil[2]{\small Institute of Earthquake Prediction Theory and Mathematical Geophysics, Russian Academy of Sciences, Moscow, Russia}
\affil[3]{\small Accademia Nazionale delle Scienze detta dei XL, Roma, Italia}
\affil[4]{\small DeepField Sensing, France}
\affil[5]{\small Muséum National d’Histoire Naturelle, CNRS, UMR 7196, INSERM U1154, Paris, France}
\affil[6]{\small Sorbonne Université, CNRS, METIS, UMR 7619, Paris, France}

\renewcommand\Authands{ and }
\makeatletter
\renewcommand\AB@affilsepx{\\[0.5em]}
\makeatother

\date{}
\maketitle

\begingroup
\renewcommand\thefootnote{\dag}
\footnotetext{\textit{In memoriam:} Jean-Louis Le Mouël passed away before the finalization of this manuscript. We dedicate this work to his memory.}
\endgroup

\begin{abstract}
	The solar cycle is commonly interpreted as an internally generated magnetic oscillation, yet its phase stability and long-term amplitude modulation remain difficult to explain through autonomous dynamo physics alone. We test the narrower hypothesis that the solar dynamo behaves as a self-excited nonlinear oscillator whose phase and envelope may be weakly modulated by coherent astronomical timing. Using historical sunspot numbers, terrestrial length-of-day records, and ephemeris-based planetary descriptors, we identify shared long-period pseudo-cycles across solar, geophysical, and astronomical observables. Singular Spectrum Analysis is used only as a non-parametric decomposition method. We then drive a reduced nonlinear $\alpha$--$\Omega$ oscillator with prescribed planetary timing signals. Solar barycentric angular momentum is treated as a mechanically defined ephemeris descriptor, whereas summed planetary right ascension is retained solely as an empirical geometrical phase proxy. We also introduce the physical planetary quadrupolar tidal tensor and use its equatorial off-diagonal component, $2T_{xy}$, as an independent timing descriptor. Forcing by normalized $2T_{xy}$ yields synchronization comparable to, and slightly better than, right ascension: $r=0.660$ and $\mathrm{RMSE}*{z}=0.824$, versus $r=0.655$ and $\mathrm{RMSE}*{z}=0.831$. Planetary forcing is not proposed to power or replace the internal dynamo, but only to provide weak timing. A possible relay involves planetary tides, tachocline magneto-Rossby dynamics, and nonlinear dynamo response. These results do not establish a complete coupling mechanism, but support the possibility that the solar cycle is susceptible to weak astronomical synchronization.

\end{abstract}

\newpage

\section{\label{sec:I}Introduction}
The solar cycle is commonly interpreted as an internally generated magnetic oscillation arising from the nonlinear dynamics of the solar dynamo. Yet, despite major progress in dynamo theory, both the phase stability of the Schwabe cycle (\cf \shortciteNP{Schwabe1844}) and the origin of its multidecadal amplitude modulation remain difficult to explain from autonomous internal dynamics alone. Over the historical interval covered by instrumental observations, sunspot activity exhibits an approximately 11-year pseudo-cycle whose amplitude is itself strongly modulated, with pronounced episodes of reduced and enhanced activity such as the Dalton Minimum and the Modern Maximum (\cf Figure \ref{Fig:01}; \eg \shortciteNP{LeMouel2010}).

This difficulty has motivated recurrent attempts to examine whether the solar cycle might behave not as a purely autonomous oscillator, but as a nonlinear self-excited system subject to weak external synchronization. In that context, planetary forcing has long been proposed as a possible candidate. As early as \shortciteN{Wolf1859}, it was suggested that the $\sim$11-year sunspot cycle could reflect the gravitational influence of Venus, Earth, Jupiter, and Saturn. Since then, the possibility of a planetary imprint on solar variability has remained controversial, chiefly because no broadly accepted physical coupling mechanism has been established and because the tidal forces exerted by the planets on the Sun appear extremely small compared with the internal forces governing the solar dynamo (\eg \shortciteNP{Callebaut2012}).

Nevertheless, the broader idea that astronomical forcing may leave measurable signatures in geophysical systems is well established. In studies of variations in the length-of-day (LOD), the Sun, the Moon, and Jupiter have long been considered as important contributors (\eg \shortciteNP{Lambeck2005,LeMouel2019a,Lopes2021}). Moreover, links between solar activity, commonly represented by the sunspot number (SSN), and LOD variations have repeatedly been reported (\eg \shortciteNP{Rosen1983,delRio2003,Lopes2017,LeMouel2019b}). These relationships are generally interpreted through changes in solar insolation, atmospheric circulation, and angular-momentum exchange between the atmosphere and the solid Earth. Yet such exchanges remain weak, while direct variations in total solar irradiance over the Schwabe cycle are themselves small, typically of order $10^{-3}$ (\eg \shortciteNP{Eddy1982,Kuhn1988,Li2012}). The question therefore remains open as to whether part of the shared variability may instead reflect a deeper dynamical connection involving common external astronomical forcing.

Several periodicities reported in solar and geophysical records are close to characteristic planetary frequencies or combinations thereof. These include timescales near $\sim$160 years (\cf \shortciteNP{Jose1965}), $\sim$90 years (\cf \shortciteNP{Gleissberg1939,leMouel2017}), $\sim$60 years, $\sim$33 years, $\sim$28 years (\eg \shortciteNP{LeMouel2023}), $\sim$22 years (\cf \shortciteNP{Hale1919}), $\sim$19 years, decadal bands around $\sim$9--14 years associated with the Schwabe (\cf \shortciteNP{Schwabe1844}) variability, as well as shorter harmonics near $\sim$5.5 and $\sim$3.6 years. Related periodicities have been described across a wide range of solar and geophysical observables (\eg \shortciteNP{Bartels1932,Zaccagnino2020,Courtillot2022,LeMouel2023,Diodato2024,Dumont2025,Xiao2025}). Their interpretation remains debated. Proposed mechanisms include internal solar variability, planetary tides, spin-orbit coupling, and the Sun's motion about the barycenter of the solar system (\eg \shortciteNP{Wood1965,Smythe1977,Fairbridge1987,Charvatova1988,Seymour1992,Hung2007,Perryman2011,Stefani2016,Cionco2018,Stefani2019,Courtillot2021,Scafetta2022,Shirley2023}).

Among these possibilities, ephemeris-based descriptors of the Sun--planet system deserve particular attention because the planetary orbital angular-momentum budget is mechanically well defined and continuously constrains the Sun's barycentric response (\eg \shortciteNP{Jose1965,Cionco2018}). We emphasize, however, that the Sun's barycentric motion is not assumed here to be a direct mechanism acting on the solar interior. It is used as a compact way to describe coherent astronomical timing and angular-momentum organization. The physical question addressed below is therefore more limited: can such weak but coherent timing modulate the phase or envelope of a nonlinear solar oscillator ?

Since then, the possibility of a planetary imprint on solar variability has remained controversial, chiefly because no broadly accepted physical coupling mechanism has been established and because the tidal forces exerted by the planets on the Sun appear extremely small compared with the internal forces governing the solar dynamo (\eg \shortciteNP{Callebaut2012}). Moreover, some direct tests have failed to identify a simple correlation between sunspot emergence geometry and planetary positions (\eg \shortciteNP{Seker2013}). These objections weigh primarily against strong or purely local triggering scenarios, but do not exclude the weaker possibility of phase or envelope modulation of a nonlinear solar oscillator.

Recent peer-reviewed literature reflects this tension. On the one hand, several studies have explored solar barycentric dynamics, tidal synchronization, nonlinear synchronization, and resonance-like interpretations of solar periodicities (\eg \shortciteNP{Cionco2018,Stefani2019,Scafetta2022,Horstmann2023,Klevs2023,Munoz2023,Stefani2023,Stefani2024}). On the other hand, critical analyses have questioned the statistical and physical basis of planetary synchronization claims, especially when formulated as direct tidal driving or strict clocking of the solar cycle (\eg \shortciteNP{Nataf2022,Weisshaar2023,Cionco2023}). This mixed literature motivates the cautious formulation adopted here: planetary motion is not treated as a source of magnetic energy, but only as a candidate weak timing signal acting on an internally sustained nonlinear oscillator.

A physically plausible weak-coupling scenario must therefore differ from direct tidal driving of the Schwabe cycle. One candidate is a magneto-Rossby relay in the tachocline. In that picture, short-period planetary spring tides provide a coherent astronomical clock, tachocline Rossby or magneto-Rossby modes provide the possible amplification and dynamical coupling, and nonlinear or noise-assisted synchronization converts the weak signal into phase or envelope modulation of the dynamo. This view is motivated by theoretical and observational studies of tachocline Rossby and magneto-Rossby dynamics, including Rieger-type and quasi-biennial periodicities, forced tachocline planetary waves, and Rossby-like magnetic activity bands (\eg \shortciteNP{Zaqarashvili2010a,Zaqarashvili2010b,Gurgenashvili2016,Gachechiladze2019,Dikpati2018,McIntosh2017,Loptien2018,Teruya2022,Horstmann2023,Korsos2023,Dikpati2024,Stefani2024,Stefani2025}). In such a framework, stochastic resonance and noise-assisted synchronization are relevant because the coherent astronomical signal need not supply the magnetic energy of the cycle; it need only bias the timing of a noisy self-excited oscillator (\eg \shortciteNP{Kramers1940,Gammaitoni1998,Hanggi1990,Albert2021,Budrikis2021}).

In this study, we revisit that question from the perspective of nonlinear synchronization. Rather than assuming that planetary forcing drives the solar cycle by itself, we test the more specific hypothesis that the solar dynamo behaves as a self-excited nonlinear oscillator that may be weakly synchronized or phase-modulated by planetary timing. To do so, we combine mechanically defined ephemeris descriptors with a data-driven analysis of key heliophysical and geophysical observations. We first examine the solar barycentric angular momentum $L_{\odot}(t)$ and its spectral content. We then compare planetary, solar, and terrestrial records using Singular Spectrum Analysis (SSA; \eg \shortciteNP{Vautard1992,Gibert2024}) in order to isolate shared pseudo-periodic components, where pseudo-periodicity denotes oscillatory modes whose phase and amplitude evolve in time. Finally, we investigate whether a reduced nonlinear $\alpha$--$\Omega$ oscillator can respond coherently to a prescribed planetary timing signal. In this last step, $RA_{\mathrm{tot}}(t)$ is used only as an empirical geometrical phase proxy; it is not interpreted as a force, torque, or unique physical coupling variable.

Our aim is therefore not to replace standard dynamo theory with an external planetary mechanism, but to assess whether weak astronomical timing provides an observationally consistent and falsifiable modulation of an otherwise internal solar magnetic cycle. The precise physical pathway by which such timing could couple to tachocline shear, differential rotation, or dynamo coefficients remains an open problem.

\section{Observational evidence for common periodic structure\label{sec:II}}

To assess whether planetary motion, solar activity, and terrestrial rotation share a common low-frequency organization, we compare four time series over the historical interval covered by instrumental or reconstructed records: the sunspot number (SSN), the length-of-day (LOD), the Sun's orbital angular momentum about the solar-system barycenter, $L_{\odot}(t)$, and a secondary empirical phase proxy derived from the summed planetary right ascensions, $RA_{\mathrm{tot}}(t)$. The first three quantities are physically distinct observables or geophysical records. The last one is not a mechanical observable; it is included only as a compact geometrical timing descriptor for testing weak synchronization in a reduced oscillator. Our objective at this stage is deliberately empirical: before introducing any causal interpretation, we ask whether these series exhibit a shared pseudo-periodic structure that is difficult to explain as isolated frequency coincidences.

\subsection{Data and observables\label{sec:II-1}}
\begin{figure}[H]
	\centering
	\includegraphics[width=\textwidth]{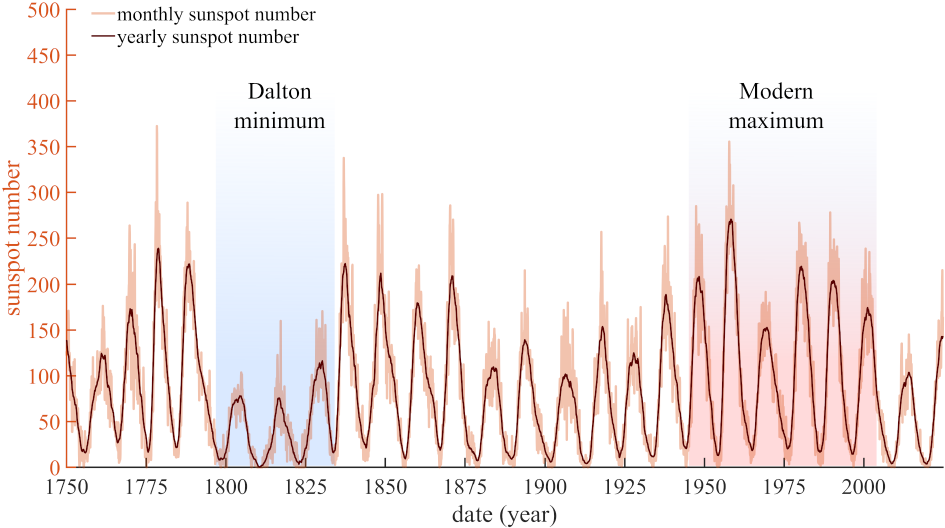}
	\caption{Evolution of the sunspot number series since 1749, shown as monthly values (in red) and annual averages (in black), as distributed by WDC-SILSO (2025). The record displays more than 24 complete Schwabe cycles  (Schwabe 1844), with characteristic periods of approximately 11 years, varying between $\sim$9 and $\sim$13 years over time (\cf Courtillot \textit{et al.} 2021). The amplitude envelope is also strongly modulated, revealing two prominent multi-decadal features: the Dalton Minimum (1800-1830; \ie low-activity shaded in blue) and the Modern Maximum (1945-2008; \ie high-activity shaded in red).}
	\label{Fig:01}
\end{figure}

Figure \ref{Fig:01} recalls the main phenomenology of the historical sunspot-number record (\cf \shortciteNP{WDCSILSO2025}). Since 1749, the SSN series displays more than 24 Schwabe cycles, with durations varying between approximately 9 and 13 years, together with a pronounced long-term amplitude modulation. Two broad envelope-scale features are particularly visible: a prolonged low-activity episode during the Dalton Minimum and a sustained high-activity interval during the Modern Maximum. Any external-timing hypothesis must therefore account not only for the existence of an approximately 11-year solar cycle, but also for the modulation of its phase and amplitude on multidecadal timescales.

To compare solar variability with planetary and terrestrial quantities, we next consider three additional time series. The primary ephemeris-based quantity is the Sun's orbital angular momentum about the solar-system barycenter, $L_{\odot}(t)$, computed from planetary ephemerides. This quantity is mechanically defined and follows from the angular-momentum budget of the Sun--planet system. We also consider a geometrical timing proxy, $RA_{\mathrm{tot}}(t)$, defined as the mass-weighted sum of the instantaneous planetary right ascensions in an inertial frame,
\begin{equation}
	RA_{\mathrm{tot}}(t)=\sum_{i=1}^{N} m_i\,RA_i(t),
\end{equation}

where $RA_i(t)$ is the heliocentric right ascension of planet $i$ at time $t$, and $m_i$ its mass. This quantity should not be read as a force, torque, or invariant mechanical coordinate. It is a compact empirical phase index of planetary configuration. Its usefulness must therefore be judged by robustness and comparison with alternative ephemeris descriptors, not by mechanical uniqueness. The third quantity is the LOD record, reconstructed from historical (\cf \shortciteNP{Stephenson1984}) observations and extended by modern geodetic measurements. We use LOD here as an independent terrestrial observable whose low-frequency variability has often been discussed in connection with external torques and angular-momentum exchanges.

\begin{figure}[H]
	\centering
	\includegraphics[width=0.7\textwidth]{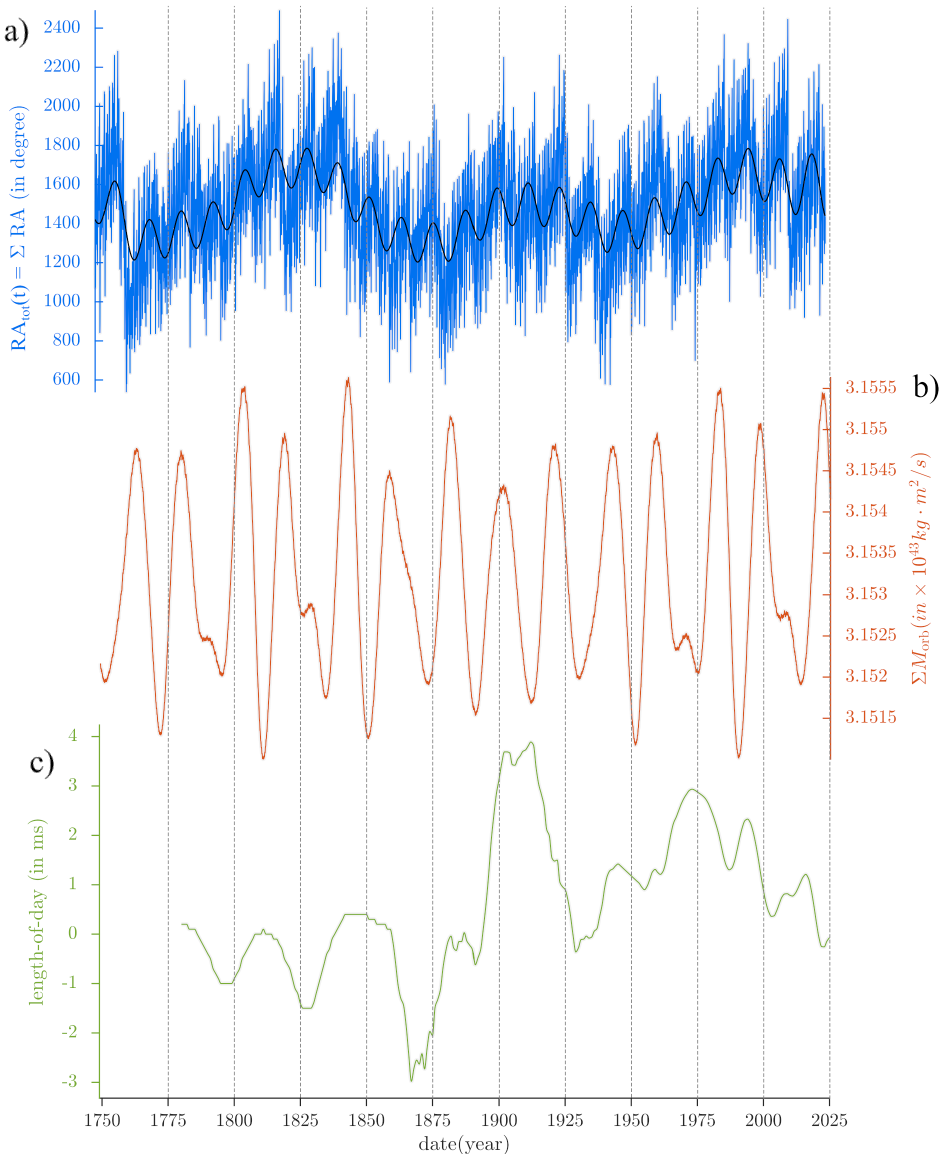}
	\caption{Temporal evolution of the three non-SSN observables considered in this study. (a) Summed planetary right ascensions, $RA_{\mathrm{tot}}(t)$; (b) Sun's orbital angular momentum about the barycenter, $L_{\odot}(t)$; (c) length of day (LOD). The three records differ strongly in waveform, but all exhibit structured variability on decadal-to-multidecadal timescales.}
	\label{fig:02}
\end{figure}

Figure \ref{fig:02} shows the temporal evolution of $RA_{\mathrm{tot}}(t)$, $L_{\odot}(t)$, and LOD over the historical interval. The three series have very different morphologies: $RA_{\mathrm{tot}}(t)$ exhibits a strongly structured saw-tooth pattern, $L_{\odot}(t)$ varies more smoothly, and LOD evolves on broader timescales with superposed shorter fluctuations. Despite these differences, all three display organized variability on decadal and longer timescales, motivating a direct comparison in the spectral and pseudo-periodic domains.

\subsection{Common spectral bands\label{sec:II-2}}
\begin{figure}[H]
	\centering
	\includegraphics[width=0.7\textwidth]{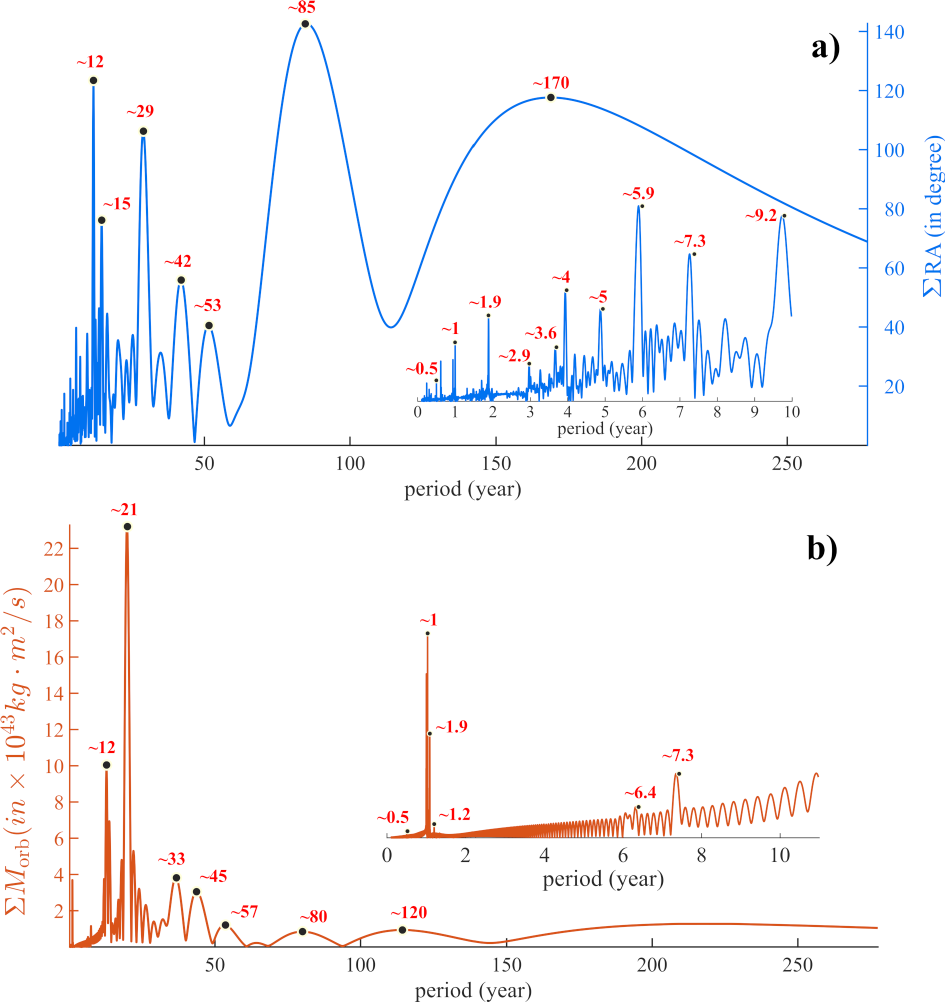}
	\caption{Fourier spectra, expressed in periods, of (a) the empirical geometrical phase proxy $RA_{\mathrm{tot}}(t)$ and (b) the mechanically defined Sun's orbital angular momentum $L_{\odot}(t)$. Common spectral bands are present in both quantities, despite different definitions and relative amplitudes. These bands overlap the principal periodicities later identified in the solar and terrestrial records.}
	\label{fig:03}
\end{figure}

A first indication of common organization is provided by the Fourier spectra of the ephemeris-based planetary descriptors. Figure \ref{fig:03} shows the period-domain spectra of $RA_{\mathrm{tot}}(t)$ and $L_{\odot}(t)$. Because these quantities have distinct definitions, and because only $L_{\odot}(t)$ is mechanically defined, the comparison is used here descriptively rather than as a causal argument. Their dominant spectral bands are nevertheless broadly consistent, including decadal, bidecadal, multidecadal, and secular-scale components that overlap the principal bands classically discussed in solar variability.

We then compare these planetary bands with those extracted from SSN and LOD. Table \ref{tab:02} summarizes the pseudo-periodic components identified in the three systems. Singular Spectrum Analysis (SSA) detects a trend and 11 oscillatory components common to SSN and LOD, while the corresponding planetary bands are present either directly in $L_{\odot}(t)$ or, for a few weaker cases, more clearly in the empirical phase proxy $RA_{\mathrm{tot}}(t)$. The resulting set spans approximately $\sim$180, $\sim$80, $\sim$55--60, $\sim$33, $\sim$27--29, $\sim$22, $\sim$19, $\sim$13--14, $\sim$11, $\sim$5.5, and $\sim$3.6 years.

\begin{table}[htbp]
	\centering
	\caption{Common pseudo-periodic components identified in SSN, LOD, and $L_{\odot}(t)$, together with the corresponding reference periods $LR^{*}$ associated with planetary orbital or Laplace-type resonance (\textit{eg} Morth \& Schlamminger 1979) timescales. Uncertainties correspond to the full width at half maximum of the spectral peak used to estimate the nominal period.}
	\label{tab:02}
	
	\vspace{0.5em}
		\renewcommand{\arraystretch}{1.25}
	\begin{tabular}{cccc}
		\toprule
		\textbf{SSN (yr)}
		& \textbf{LOD (yr)}
		& \textbf{$L_{\odot}$ (yr)}
		& \textbf{$LR^{*}$ ref.} \\
		\midrule
		
		trend & trend & -- & -- \\
		
		$\sim 182.42$
		& $\sim 186.2$
		& $200 \pm 45$
		& [1] \\
		
		$77.71 \pm 13.63$
		& $83.25 \pm 17.83$
		& $81.36 \pm 5.47$
		& [2] \\
		
		$54.38 \pm 16.20$
		& $50.36 \pm 9.09$
		& $56.83 \pm 2.11$
		& [3] \\
		
		$34.35 \pm 2.79$
		& $32.63 \pm 2.84$
		& $33.15 \pm 1.58$
		& [4] \\
		
		$26.81 \pm 2.04$
		& $27.60 \pm 3.33$
		& $28.95 \pm 1.26$
		& [5] \\
		
		$21.25 \pm 1.03$
		& $22.27 \pm 1.30$
		& $21.06 \pm 1.02$
		& [6] \\
		
		$18.23 \pm 1.09$
		& $19.37 \pm 1.11$
		& $20.03 \pm 0.33$
		& [7] \\
		
		$13.12 \pm 0.38$
		& $13.26 \pm 0.44$
		& $14.15 \pm 0.23$
		& [8] \\
		
		$11.01 \pm 0.33$
		& $10.45 \pm 0.37$
		& $11.92 \pm 0.25$
		& [9] \\
		
		$5.48 \pm 0.07$
		& $5.34 \pm 0.17$
		& $5.81 \pm 0.12$
		& [10] \\
		
		$3.88 \pm 0.03$
		& $3.61 \pm 0.03$
		& $3.60 \pm 0.10$
		& [11] \\
		
		\bottomrule
	\end{tabular}
	
	\vspace{0.8em}
	
	\begin{minipage}{0.96\linewidth}
		\footnotesize
		$^{[1]}$ de Vries / Neptune (\textbf{164.89});
		$^{[2]}$ Gleissberg / Uranus (\textbf{84.01});
		$^{[3]}$ Neptune + Uranus (\textbf{57.29});
		$^{[4]}$ Jupiter + Uranus (\textbf{36.06});
		$^{[5]}$ Jupiter + Saturn + Uranus (\textbf{25.57});
		$^{[6]}$ Hale / Jupiter + Saturn (\textbf{21.64});
		$^{[7]}$ Jupiter + Saturn (\textbf{18.61});
		$^{[8]}$ Schwabe band / Jupiter (\textbf{11.85});
		$^{[9]}$ Schwabe band / Jupiter--Saturn (\textbf{9.31});
		$^{[10]}$ Jupiter (\textbf{5.92});
		$^{[11]}$ Jupiter (\textbf{3.91}).
	\end{minipage}
	
\end{table}

At this stage, the key empirical point is not that every period matches a single unique planetary frequency exactly, but that the same ranked set of decadal-to-multidecadal bands recurs independently across planetary, solar, and terrestrial observables. This common structure already constrains any acceptable interpretation.

Table~\ref{tab:02} should not be read as an exhaustive peak-picking list of the raw $L_\odot(t)$ spectrum. Its purpose is more limited: the first two columns define the ranked pseudo-periodic bands common to SSN and LOD, while the planetary columns indicate nearby ephemeris-based bands that may provide coherent astronomical timing. Consequently, peaks visible in the raw $L_\odot(t)$ spectrum but not belonging to the common SSN--LOD sequence, such as the $\sim 120$ yr and $\sim 45$ yr features or the short-period peaks near $1$--$2$ yr, are not used in the statistical comparison summarized in Figure~\ref{fig:06}. Similarly, the $\sim 21$ yr entry should be understood as a broad Hale-scale or bidecadal band, not as a strict one-to-one identification with a unique barycentric period. The comparison therefore concerns the organization of period bands, not a direct mechanical assignment of every spectral maximum of $L_\odot(t)$ to a solar or terrestrial mode.

\subsection{Shared pseudo-periodic modes\label{sec:II-3}}
Spectral agreement alone does not establish whether the corresponding modes evolve similarly in time. We therefore use SSA to extract the low-frequency trend and the 11 common pseudo-periodic components from SSN and LOD. Here, pseudo-periodicity denotes oscillatory modes whose phase and amplitude are allowed to vary slowly with time. The purpose of this decomposition is descriptive: it isolates coherent low-rank structures in the data and allows their temporal evolution to be compared directly.

\begin{figure}[H]
	\centering
	\includegraphics[width=0.8\textwidth]{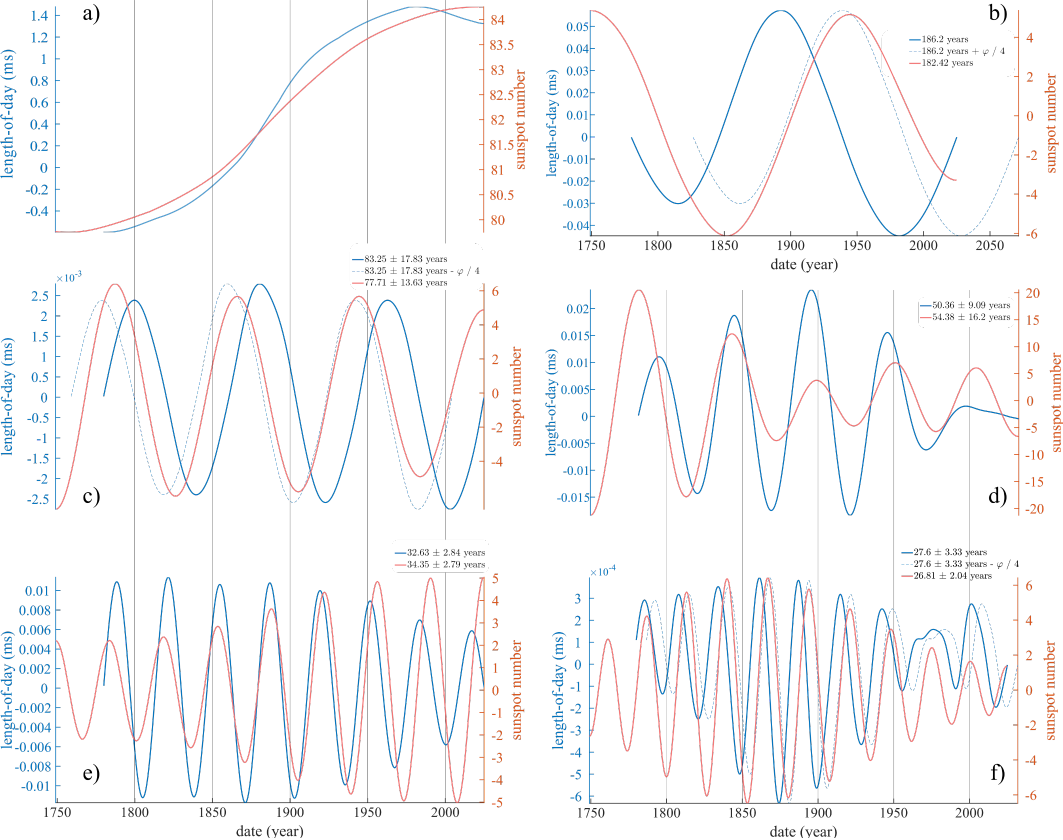}
	\caption{SSA-reconstructed long-period components of SSN (red) and LOD (blue): (a) trend, (b) $\sim$180-year band, (c) $\sim$80-year band, (d) $\sim$55--60-year band, (e) $\sim$33-year band, and (f) $\sim$27--29-year band. The dominant low-frequency modes show substantial coherence in envelope and broad phase organization.}
	\label{fig:04}
\end{figure}

\begin{figure}[H]
	\centering
	\includegraphics[width=0.8\textwidth]{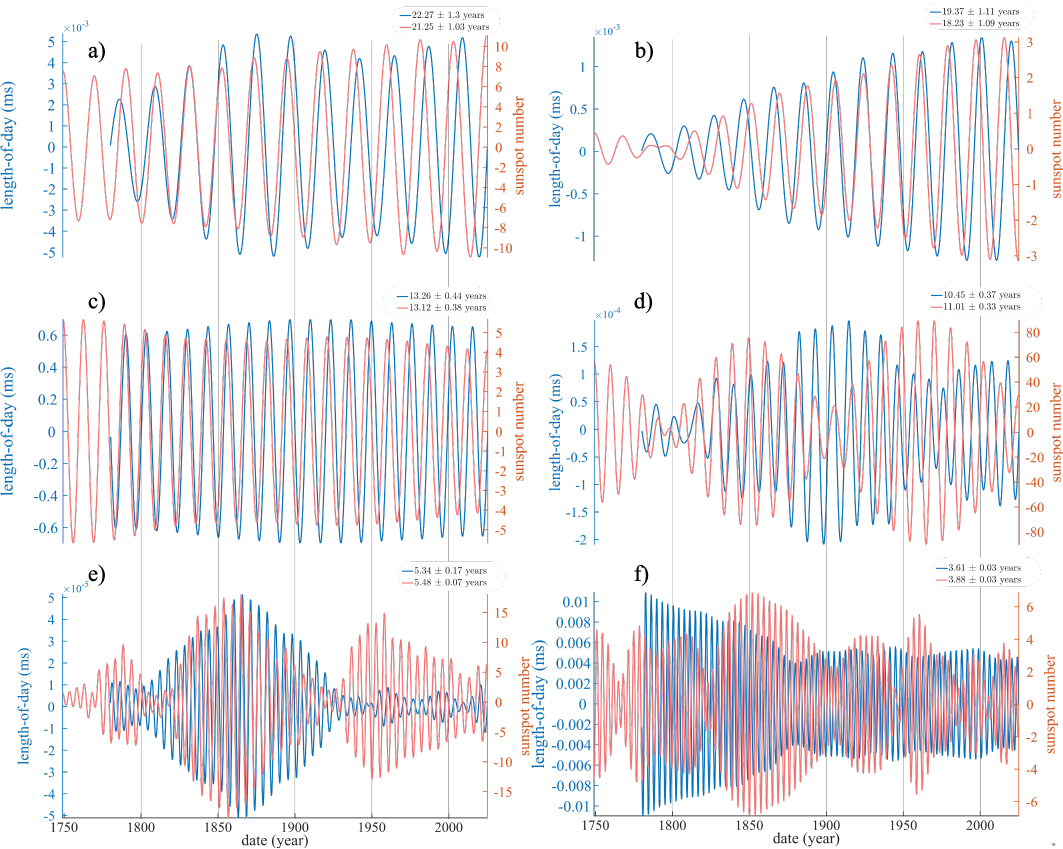}
	\caption{SSA-reconstructed decadal components of SSN (red) and LOD (blue): (a) $\sim$22-year band, (b) $\sim$19-year band, (c) $\sim$13--14-year band, (d) $\sim$11-year band, (e) $\sim$5.5-year band, and (f) $\sim$3.6-year band. The principal solar-cycle bands and their shorter harmonics are present in both records.}
	\label{fig:05}
\end{figure}

Figures \ref{fig:04} and \ref{fig:05} show the reconstructed trend and the 11 pseudo-periodic components, ordered from the longest to the shortest period. Several robust features emerge. First, the longest modes, including the secular trend, the $\sim$180-year band, and the $\sim$80-year band, show strongly coherent envelopes and, in some cases, near-quadrature phase relationships. Second, the intermediate bands around $\sim$60, $\sim$33, and $\sim$27$-$29 years remain similar in overall structure, although modest phase drift becomes more visible. Third, the decadal bands associated with the Schwabe/Hale range show clear similarity in both amplitude modulation and recurrence structure. Finally, the shorter $\sim$5.5-year and $\sim$3.6-year modes remain detectable in both SSN and LOD, even though their amplitudes and long-term envelope behavior are less regular.

Taken together, these reconstructions show that the common spectral bands are not merely isolated frequency coincidences. They correspond to structured pseudo-periodic modes whose time-dependent envelopes and relative phases remain comparable over substantial parts of the historical record. In other words, the shared organization extends from the Fourier domain into the temporal domain.

\subsection{Statistical support against chance agreement\label{sec:II-4}}
To quantify how unusual this agreement is, we compare the extracted periods statistically. The correlation between the 11 detected periods in SSN and LOD is close to unity. Bootstrap (\cf \shortciteNP{Efron1994}) tests indicate that such a high correspondence is unlikely under the simple ordered-random null model used in Appendix~\ref{app:A}. This test is intentionally conservative in interpretation: it is a diagnostic against accidental ranked matching, not a definitive proof of common origin. We further compare the empirical cumulative distributions of the detected periods in SSN, LOD, and $L_{\odot}(t)$ with the reference distribution $LR^{*}$ associated with the planetary resonance-like timescales listed in Table~\ref{tab:02}.

\begin{figure}[H]
	\centering
	\includegraphics[width=0.8\textwidth]{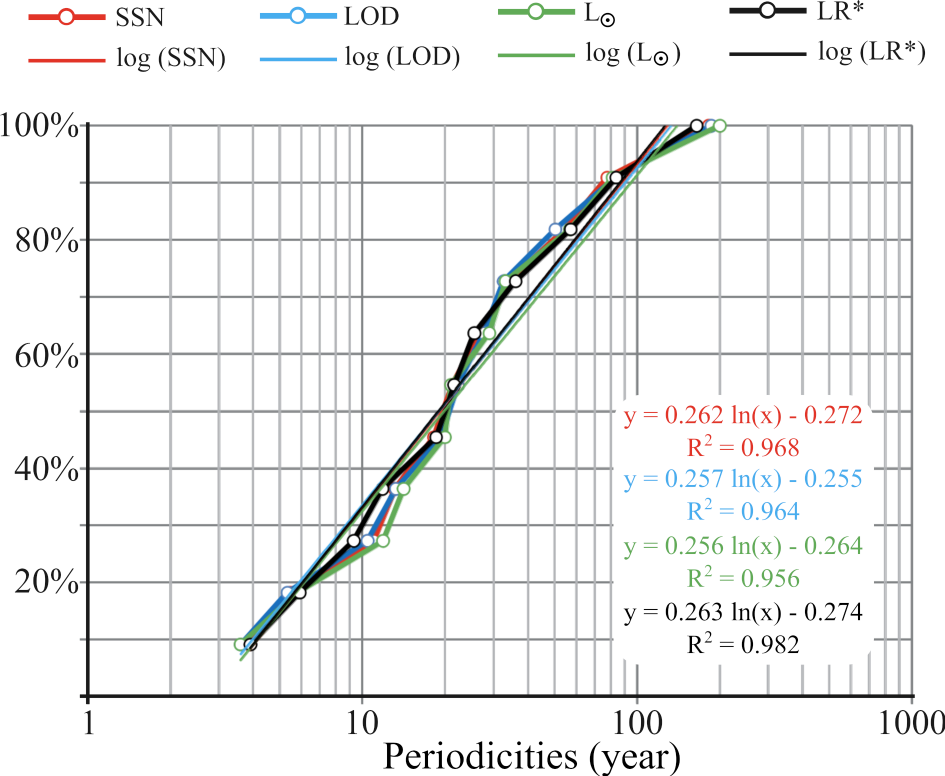}
	\caption{Empirical cumulative distribution functions of the detected pseudo-periods for SSN (red), LOD (blue), and $L_{\odot}(t)$ (green), compared with the reference distribution $LR^{*}$ (black). The close alignment of the four curves indicates that the ranked periodic structures are highly similar across the three observables and the planetary reference set.}
	\label{fig:06}
\end{figure}

Figure \ref{fig:06} shows that the empirical cumulative distributions of SSN, LOD, and $L_{\odot}(t)$ closely track the reference distribution. This agreement provides an independent summary of the spectral results: the three observables do not merely contain a few overlapping bands, but share a similar ranked periodic organization. We therefore regard the common decadal-to-multidecadal structure identified in SSN, LOD, and planetary observables as unlikely to be explained by isolated coincidences alone under the null models considered here. This empirical result does not specify the mechanism, and stronger tests using red-noise surrogates and full-pipeline SSA resampling remain desirable. It nevertheless defines the central dynamical puzzle addressed in the remainder of this study: why should solar, terrestrial, and planetary records display related low-frequency pseudo-periodic architectures ?

\section{Ephemeris-based timing and weak-forcing interpretation\label{sec:III}}
{The empirical results of Section~\ref{sec:II} show that planetary, solar, and terrestrial records share a similar low-frequency pseudo-periodic structure. This observation does not by itself identify a mechanism, but it motivates a limited and testable question: can a weak and coherent astronomical timing signal modulate an otherwise internal solar oscillator ? In this section, we introduce two complementary families of ephemeris-based descriptors. The first is barycentric and angular-momentum based; the second is tidal and local in the sense that it describes the quadrupolar gravitational field imposed by the planets at the Sun. Neither family is assumed to power the solar cycle. Both are used to construct physically cautious timing descriptors for weak-synchronization tests.

\subsection{Barycentric frame and solar orbital motion\label{sec:III-1}}
We consider the solar system in a Galilean barycentric reference frame. The barycenter is the center of mass of the Sun--planet system and is defined by
\begin{equation}
	\mathbf{R}_{b}=\frac{\sum_{i=1}^{N} m_i \mathbf{r}_i}{\sum_{i=1}^{N} m_i},
	\label{eq:02}
\end{equation}

where $m_i$ and $\mathbf{r}_i$ denote the mass and position of body $i$, respectively. Because the Sun ($M_{\odot}$) dominates the total mass budget, the barycenter remains close to the solar interior, but it is not fixed at the Sun's geometric center. Its location changes continuously with planetary configuration, chiefly under the influence of the giant planets.

In this frame, the barycenter moves at constant velocity in the absence of external forces, while the Sun itself describes a small orbit around it. The corresponding barycentric trajectory is shown in Figure~\ref{fig:07}, computed from planetary ephemerides. Over the historical interval considered here, the barycenter may lie either inside or outside the solar radius, and the displacement of the Sun relative to that barycenter reaches values of order a few solar radii. This motion is entirely Newtonian and follows directly from the gravitational interaction of the Sun with the planets.
\begin{figure}[H]
	\centering
	\includegraphics[width=0.8\textwidth]{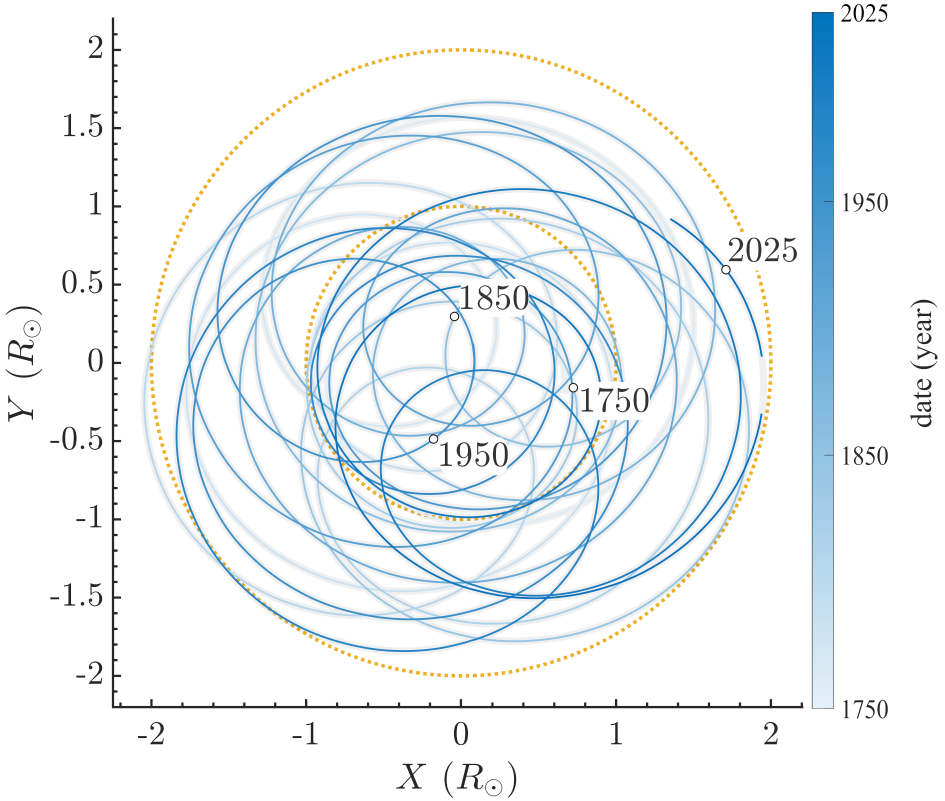}
	\caption{Temporal evolution of the solar-system barycenter relative to the Sun from 1750 to 2025, computed in a Galilean barycentric reference frame from planetary ephemerides. The trajectory is shown in the $XY$ plane and expressed in solar radii, with color evolving from light blue (1750) to dark blue (2025). Depending on planetary configuration, the barycenter lies either inside or outside the solar radius and spans a displacement of order a few $R_{\odot}$.}
	\label{fig:07}
\end{figure}

The Sun's barycentric motion obeys the same Newtonian principles as planetary motion. Writing $\mathbf{r}_{\odot}$ for the Sun's barycentric position, its acceleration satisfies,
\begin{equation}
	M_{\odot}\ddot{\mathbf{r}}_{\odot}
	=
	-\sum_{j}\mathcal{G}M_{\odot}m_j
	\frac{\mathbf{r}_{\odot}-\mathbf{r}_j}{\lVert \mathbf{r}_{\odot}-\mathbf{r}_j\rVert^3},
	\label{eq:03}
\end{equation}

where the sum runs over the planets. This equation simply expresses that the Sun responds to the same gravitational field generated by the planetary system that simultaneously governs planetary motion itself.

\subsection{Angular-momentum balance\label{sec:III-2}}
The quantity of primary interest for the barycentric part of the analysis is the angular-momentum budget of the Sun--planet system. In a closed barycentric frame, the total angular momentum may be written as the sum of orbital and rotational contributions,
\begin{equation}
	M_{\mathrm{tot}}
	=
	\sum_{i=1}^{N}\left(m_i\,\mathbf{r}_i\times\mathbf{v}_i\right)
	+
	I_i\boldsymbol{\Omega}_i,
	\label{eq:01b}
\end{equation}

where $I_i\boldsymbol{\Omega}_i$ denotes the spin angular momentum of body $i$. In the absence of external torque, the total angular momentum is conserved.

Although the Sun contains nearly all of the mass of the solar system, the orbital angular-momentum budget is dominated by the Jovian planets, especially Jupiter and Saturn. As planetary configurations evolve, the orbital angular momentum carried by the planets changes continuously, and this variability is compensated within the total conserved barycentric budget. A natural observable is therefore the Sun's orbital angular momentum about the barycenter,
\begin{equation}
	\underbrace{M_{\odot}\mathbf{r}_{\odot}\times\mathbf{v}_{\odot}}_{\mathbf{L}_{\odot}}
	+
	\underbrace{\sum_i m_i\,\mathbf{r}_i\times\mathbf{v}_i}_{M_{\mathrm{tot}}-\sum_i I_i\boldsymbol{\Omega}_i}
	=
	\mathbf{C},
	\label{eq:04}
\end{equation}

where $\mathbf{C}$ is the conserved barycentric angular-momentum vector.

After subtracting the constant part of the barycentric budget, one may focus on fluctuations around the mean and write, to very good approximation,
\begin{equation}
	\mathbf{L}_{\odot}(t)\approx -\sum_i \mathbf{L}_i(t),
	\label{eq:05}
\end{equation}

that is, the Sun's barycentric orbital angular momentum varies oppositely to the summed planetary orbital contribution. In this approximation, the planetary spin terms are neglected because they are several orders of magnitude smaller than the dominant orbital terms. This is the reason why $L_{\odot}(t)$ provides a compact and dynamically meaningful measure of the planetary contribution to the solar barycentric motion.

The quantity $RA_{\mathrm{tot}}(t)$ introduced in Section~\ref{sec:II} plays a more limited role. Whereas $L_{\odot}(t)$ is mechanically defined, $RA_{\mathrm{tot}}(t)$ only tracks the evolving relative orientation of the planets in a compact empirical form. It is therefore used below as one possible phase-bearing timing proxy, not as a dynamically invariant observable. The distinction is important: $L_{\odot}(t)$ constrains the barycentric angular-momentum organization, whereas $RA_{\mathrm{tot}}(t)$ is a geometrical descriptor whose robustness must be tested against physically defined alternatives.

\subsection{Planetary tidal tensor and magneto-Rossby relay\label{sec:III-3}}
The geometrical character of $RA_{\mathrm{tot}}(t)$ motivates the use of a physically defined tidal descriptor. For a planet of mass $m_i$, heliocentric distance $r_i$, and unit direction vector $\mathbf{n}_i$, the quadrupolar tidal tensor at the Sun is,
\begin{equation}
	T_{ab}(t)=\sum_i \frac{\mathcal{G}m_i}{r_i^3(t)}\left[3n_{i,a}(t)n_{i,b}(t)-\delta_{ab}\right].
	\label{eq:tidal_tensor}
\end{equation}

In the equatorial plane, the complex quadrupolar component may be written as,
\begin{equation}
	Z_2(t)=\left[T_{xx}(t)-T_{yy}(t)\right]+2iT_{xy}(t),
	\label{eq:tidal_z2}
\end{equation}

so that $2T_{xy}(t)$ is a genuine off-diagonal component of the planetary tidal tensor. Unlike $RA_{\mathrm{tot}}(t)$, it depends on planetary masses, distances, and directions through the physical scaling $m_i/r_i^3$. We therefore use $2T_{xy}(t)$ as the main physically defined tidal timing descriptor in the dynamical test below, while retaining $RA_{\mathrm{tot}}(t)$ as an empirical phase-bearing comparison.

This construction does not remove the energetic objection to direct planetary driving; rather, it places the test in the weak-forcing regime. The dominant planetary tides are far too small to power the solar cycle directly. A viable pathway must instead act as a relay. The most plausible entry channel is the short-period planetary spring-tide spectrum, with periods of order $100$--$300$ days, which overlaps the broad range of Rossby and magneto-Rossby periods discussed for tachocline dynamics (\eg \shortciteNP{Zaqarashvili2010a,Zaqarashvili2010b,Gurgenashvili2016,Gachechiladze2019,Teruya2022,Horstmann2023,Korsos2023,Stefani2024,Stefani2025}). The tachocline response may then modulate shear, flow, or effective dynamo coefficients, while the nonlinear dynamo can rectify or beat these shorter signals into Schwabe-scale phase or envelope organization. In this interpretation, $2T_{xy}(t)$ is not the energy source of the solar cycle. It is a physically defined tidal timing descriptor that can be used to test whether the empirical timing represented by $RA_{\mathrm{tot}}(t)$ survives in a genuine tidal quantity.

\subsection{Why weak forcing is physically admissible\label{sec:III-4}}
A recurrent objection to planetary-solar coupling is that planetary forces are extremely small compared with the characteristic forces acting in the solar interior. That objection is decisive against any hypothesis of strong direct driving, but it is less restrictive in the context considered here. The question is not whether planetary forcing can power the solar cycle, but whether a weak, coherent, and long-lasting external modulation can influence the phase or envelope of a nonlinear self-excited oscillator.

As a minimal consistency check, we recall that the Sun--planet system is a gravitationally bound system governed by classical mechanics. For a Newtonian potential of the form $1/r$, the Virial theorem gives,
\begin{equation}
	2\langle T\rangle = -\langle U\rangle,
	\qquad
	\langle E\rangle = \langle T\rangle + \langle U\rangle = -\langle T\rangle < 0,
	\label{eq:01a}
\end{equation}

so that the long-term mean total energy remains negative and the motion stays bounded. In the present context, this relation is not a solar-dynamo mechanism. It is invoked only to emphasize that the ephemeris descriptors used here belong to a bounded Newtonian many-body system.

This point matters because a weak modulation of angular momentum or rotation rate need not dominate the solar energy budget in order to affect a nonlinear oscillator. In many nonlinear systems, a small coherent forcing can influence timing, phase locking, or long-term envelope modulation without replacing the internal mechanism that sustains the oscillation. The empirical results of Section~\ref{sec:II} motivate precisely this type of interpretation: planetary forcing, if relevant, should be understood as a weak timing modulation superimposed on an internally generated solar cycle, not as an alternative energy source for the dynamo.

This weak-forcing perspective is also consistent with broader nonlinear-dynamo literature. Stochastic resonance can allow weak external modulation to organize transitions in a noisy dynamical system (\eg \shortciteNP{Albert2021}), while global simulations have shown that solar-like dynamos can exhibit bistability and distinct cycle regimes (\eg \shortciteNP{Matilsky2020}). More directly, synchronized low-order or reduced solar-dynamo models have recently been discussed in both two-dimensional and nonlinear-dynamics settings (\eg \shortciteNP{Klevs2023,Munoz2023}). The relevant conceptual comparison is therefore not between planetary work and the total magnetic energy of the solar cycle, but between a relay-amplified coherent modulation and the effective noise or phase diffusion of a nonlinear dynamo variable (\eg \shortciteNP{Kramers1940,Gammaitoni1998,Hanggi1990,Budrikis2021,Albert2021}). These studies do not establish the present interpretation, but they show that weak synchronization is the relevant dynamical language, rather than strong energetic driving.

In that perspective, the barycentric and tidal quantities introduced above provide mechanically constrained timing descriptors, not a complete coupling theory. The quantity $L_{\odot}(t)$ summarizes the Sun's orbital response to the planetary angular-momentum budget. The quantity $RA_{\mathrm{tot}}(t)$ captures one aspect of the evolving planetary configuration and is used only as an empirical phase proxy. The quantity $2T_{xy}(t)$ provides a physically defined quadrupolar tidal descriptor of the same planetary configuration. This motivates the next step of the analysis: testing whether a reduced nonlinear solar oscillator can respond coherently to weak prescribed astronomical timing signals.

\section{Reduced nonlinear dynamo under prescribed astronomical timing\label{sec:IV}}
The observational results of Section~\ref{sec:II} and the weak-forcing interpretation of Section~\ref{sec:III} motivate a dynamical test: can a minimal nonlinear solar oscillator respond coherently to prescribed astronomical timing signals and reproduce the principal decadal and multidecadal features of the historical sunspot record ? We address this question using a reduced $\alpha$--$\Omega$ dynamo model written as a system of two coupled nonlinear ordinary differential equations for the temporal amplitudes of the toroidal and poloidal magnetic fields. The model is a synchronization experiment, not a complete physical model of the solar interior. We compare two inputs: the empirical phase proxy $RA_{\mathrm{tot}}(t)$ and the physically defined tidal-tensor component $2T_{xy}(t)$.

\subsection{A minimal forced $\alpha$--$\Omega$ dynamo\label{sec:IV-1}}
We adopt a low-order dynamical system corresponding to the leading temporal mode of the mean-field solar dynamo (\eg \shortciteNP{Parker1955,Weiss1993,Tobias1995,Knobloch1996,Charbonneau2023}). In the full mean-field formulation, the poloidal field $A$ and toroidal field $B$ depend on space and time, and their evolution includes differential rotation, turbulent induction, magnetic diffusion, transport, and nonlinear quenching. When projected onto the dominant dipolar mode, the spatial dependence is absorbed into effective coefficients, yielding a reduced temporal model that retains the core dynamical feedback between the $\alpha$-effect and the $\Omega$-effect.

In this reduced description, linear damping terms represent diffusive losses, while cubic terms account phenomenologically for magnetic saturation and flux-loss effects. External forcing is introduced through a normalized prescribed timing signal $X_p(t)$. The signal $X_p(t)$ is taken either from the empirical geometrical phase proxy $RA_{\mathrm{tot}}(t)$ or from the normalized physical tidal descriptor $2T_{xy}(t)$. Both signals are used only as prescribed weak timing perturbations of the reduced oscillator; neither is interpreted as a mechanical force acting on the solar plasma, nor as an energy source for the magnetic cycle. The forced dynamo is written as,
\begin{equation}
	\begin{aligned}
		\frac{dB}{dt}
		&=
		\Omega_0 A(t)
		-\eta_B B(t)
		-\mu_B B^3(t)
		+F_B(t), \\
		\frac{dA}{dt}
		&=
		\alpha_0 B(t)
		-\eta_A A(t)
		-\mu_A A^3(t)
		+F_A(t),
	\end{aligned}
	\label{eq:dynamo}
\end{equation}

where $A(t)$ and $B(t)$ denote the large-scale poloidal and toroidal amplitudes, respectively, $\eta_{A,B}$ are effective damping coefficients, and $\mu_{A,B}$ parameterize nonlinear saturation. The term $F_B(t)$ denotes the normalized timing input derived either from $RA_{\mathrm{tot}}(t)$ or from $2T_{xy}(t)$, while $F_A(t)$ represents the auxiliary modulation retained from the reference reduced-dynamo formulation. These forcing terms should therefore be read phenomenologically: they do not represent the energetic work of the planets on the solar magnetic field, but an effective weak timing perturbation acting on the phase and envelope of the self-excited oscillator.

The numerical values used for the integrations shown in Figure~\ref{fig:10} are listed explicitly here to make the reduced experiment reproducible. We used,
\begin{equation*}
	\Omega_0 = 0.60,\qquad
	\alpha_0 = 0.54,\qquad
	\eta_A = 0.12,\qquad
	\eta_B = 0.10,\qquad
	\mu_A = 0.08,\qquad
	\mu_B = 0.20 .
\end{equation*}

The corresponding intrinsic damped natural period and the one-at-a-time robustness check of the (2T$_{xy}$)-forced solution are reported in Appendix \ref{app:B}

This interpretation is central and consistent with recent work showing that tidally forced large-scale waves may arise in the tachocline of solar-like stars and can, in principle, provide a pathway for weak external modulation of dynamo-relevant shear (\eg \shortciteNP{Horstmann2023}). It is also consistent with the broader magneto-Rossby picture in which short-period tidal forcing is processed by tachocline waves and nonlinear beating before it appears as a lower-frequency modulation of a dynamo-relevant variable (\eg \shortciteNP{Zaqarashvili2010a,Gurgenashvili2016,Gachechiladze2019,Dikpati2024,Stefani2025}). The model is not constructed to replace the internal solar dynamo by an externally driven mechanism. Instead, it tests whether a self-excited nonlinear oscillator operating near its natural decadal timescale can have its phase and long-term envelope modulated by a weak, coherent, and astronomically prescribed forcing. In that sense, equation~(\ref{eq:dynamo}) should be viewed as a minimal synchronization model rather than as a complete solar-dynamo theory.

\subsection{From an empirical phase proxy to a tidal-tensor timing descriptor\label{sec:IV-2}}
Before integrating the forced dynamo, it is useful to compare the empirical planetary phase proxy directly with the sunspot record. Figure~\ref{fig:08} shows the band-limited component of $RA_{\mathrm{tot}}(t)$ in the Schwabe range together with the observed SSN, as well as their slowly varying phase difference. The upper panel shows that the proxy and the solar cycle share comparable decadal pacing, while the lower panel indicates that their relative phase evolves gradually over the historical interval rather than fluctuating randomly. This does not make $RA_{\mathrm{tot}}(t)$ a mechanical forcing variable. It only shows that this empirical phase descriptor carries coherent timing information that can be used in a weak-synchronization test. The slow phase drift visible across the record is also consistent with modulation rather than rigid locking. The corresponding physical descriptor used below is $2T_{xy}(t)$, the off-diagonal equatorial component of the planetary quadrupolar tidal tensor.

\begin{figure}[H]
	\centering
	\includegraphics[width=\textwidth]{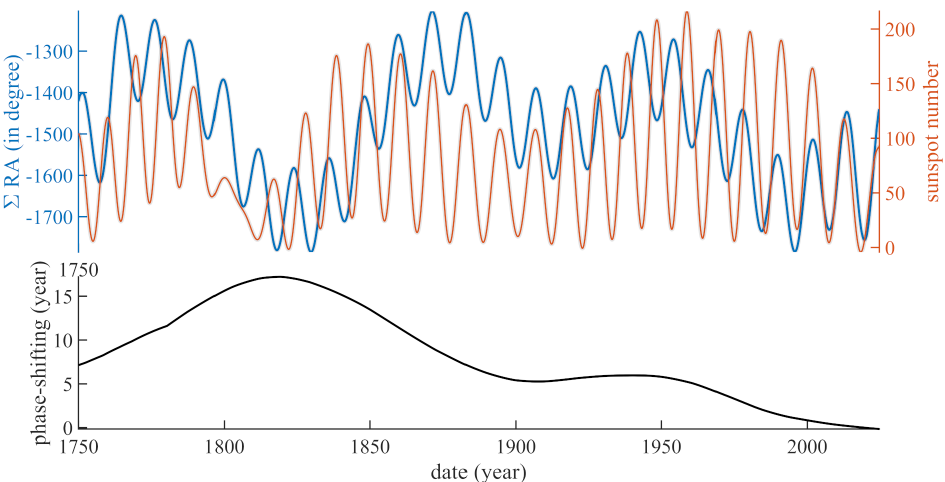}
	\caption{Top panel: comparison between the band-limited empirical phase proxy $RA_{\mathrm{tot}}(t)$ (blue, polarity inverted) and the observed sunspot number series (red). Bottom panel: slowly varying mean phase offset between the two band-limited Schwabe-range signals. This curve is a smoothed phase diagnostic and should not be interpreted as a cycle-by-cycle delay between individual maxima.}
	\label{fig:08}
\end{figure}

Within this framework, the sign convention applied to $RA_{\mathrm{tot}}(t)$ is not physically restrictive because $RA_{\mathrm{tot}}(t)$ is used only after normalization as a prescribed timing perturbation in the reduced oscillator. The effective sign of this perturbation can be absorbed into the coupling convention of the phenomenological term $F_B(t)$ and does not alter the interpretation of the model as a weakly forced nonlinear oscillator. What matters for this minimal test is not the arbitrary polarity of the empirical proxy, but the temporal organization it carries.

Figure~\ref{fig:txy_evolution} shows the temporal evolution of the physical tidal-tensor descriptor $2T_{xy}(t)$. The signal is rapidly varying because it contains the short-period tidal spectrum of the planetary system. In the dynamo experiment, we use the detrended and normalized raw $2T_{xy}(t)$ signal, rather than the smoothed envelope, so that the test is based on the physical tidal descriptor itself.

\begin{figure}[H]
	\centering
	\includegraphics[width=\textwidth]{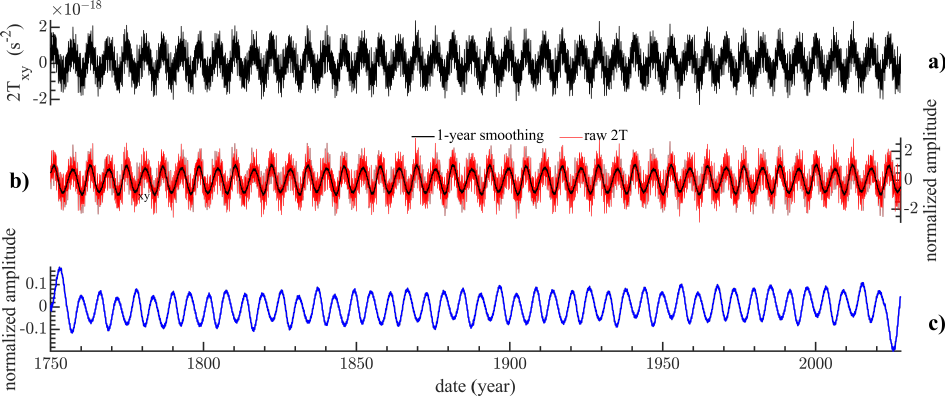}
	\caption{Temporal evolution of the physical quadrupolar tidal-tensor component $2T_{xy}(t)$ computed from planetary masses, heliocentric distances, and directions. (a) Raw $2T_{xy}(t)$ signal in SI units. (b) Detrended and normalized $2T_{xy}(t)$ signal, together with its 2 yr smoothed version for visual clarity. (c) Smoothed low-frequency modulation of the normalized tidal descriptor, shown only as a guide to the eye.}
	\label{fig:txy_evolution}
\end{figure}

This sequence of diagnostics defines the logic of the dynamical test below. The empirical proxy $RA_{\mathrm{tot}}(t)$ is first used to identify coherent planetary timing in the Schwabe band, but it is not assigned any mechanical status. The tidal component $2T_{xy}(t)$ is then introduced as a physically defined descriptor of the same planetary configuration. The key question is therefore whether the reduced dynamo response obtained with $RA_{\mathrm{tot}}(t)$ persists when this empirical proxy is replaced by the tidal-tensor forcing $2T_{xy}(t)$.

\subsection{Reproduction of the historical solar-cycle envelope\label{sec:IV-3}}
Figure~\ref{fig:10} provides the dynamical test corresponding to the timing analysis of the previous subsections. The same reduced nonlinear $\alpha$--$\Omega$ dynamo is integrated twice, using two prescribed astronomical inputs for the toroidal timing perturbation $F_B(t)$. In the first case, the input is the empirical geometrical proxy $RA_{\mathrm{tot}}(t)$ introduced above. In the second case, it is replaced by the physically defined quadrupolar tidal-tensor component $2T_{xy}(t)$. The model output shown in the figure is the toroidal component $B(t)$, smoothed over 2 yr for visualization, and compared with the reconstructed sunspot number series.

\begin{figure}[H]
	\centering
	\includegraphics[width=\textwidth]{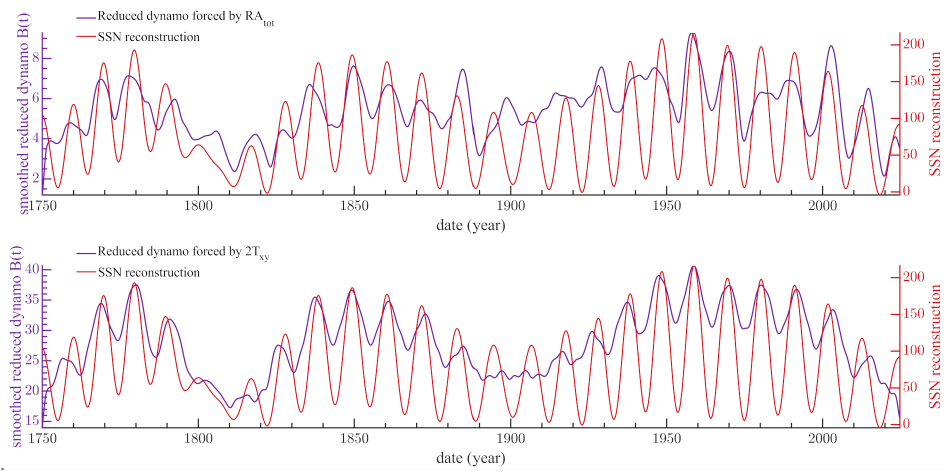}
	\caption{Reproduction of the historical solar-cycle envelope by the reduced nonlinear dynamo under two prescribed astronomical timing inputs. The violet curves show the smoothed toroidal dynamo output $B(t)$, and the red curves show the reconstructed sunspot number series. Top panel: dynamo forced by the empirical phase proxy $RA_{\mathrm{tot}}(t)$. Bottom panel: dynamo forced by the physical quadrupolar tidal-tensor component $2T_{xy}(t)$. The two vertical axes are different because the reduced dynamo output is expressed in relative units and is compared to the SSN reconstruction only through its timing, envelope, and normalized statistical agreement. The reported values of $r$ and $\mathrm{RMSE}_{z}$ quantify the correlation and normalized error between the model output and the SSN reconstruction.}
	\label{fig:10}
\end{figure}

The upper panel shows that the historical $RA_{\mathrm{tot}}(t)$ proxy remains an effective empirical timing signal for the reduced dynamo. The model reproduces the broad organization of the reconstructed sunspot record, including Schwabe-scale recurrence, the low activity interval around the Dalton-minimum epoch, the recovery during the nineteenth century, the enhanced twentieth-century envelope, and the decline toward the most recent cycles. Quantitatively, this run gives $r=0.655$ and $\mathrm{RMSE}_{z}=0.831$. This agreement confirms that the phase information identified in Section~\ref{sec:IV-2} can be transferred into a nonlinear oscillator and converted into a solar-like magnetic response. However, $RA_{\mathrm{tot}}(t)$ remains an empirical angular descriptor; it should not be interpreted as a force, torque, or energy source.
	
The lower panel is therefore the more important comparison. When the empirical angular proxy is replaced by the physical tidal-tensor component $2T_{xy}(t)$, the same reduced dynamo gives a response that is at least as good, with $r=0.660$ and $\mathrm{RMSE}_{z}=0.824$. The improvement is modest in absolute numerical terms, but it is conceptually significant: the synchronization does not depend uniquely on a geometrical sum of planetary right ascensions. It can also be obtained from an explicitly mechanical descriptor of the planetary quadrupolar tidal field. In particular, the $2T_{xy}$-forced solution follows the long-term amplitude envelope of the SSN reconstruction across the main historical intervals, while preserving the decadal pacing of the oscillations.
	
The relevant observable in Figure~\ref{fig:10} is not the absolute amplitude of $B(t)$. The reduced dynamo is a low-order oscillator, and its toroidal component is expressed in arbitrary relative units. The meaningful comparison is instead based on the temporal organization of the solution: the persistence of Schwabe-like cycles, the modulation of their envelope, and the normalized agreement with the reconstructed SSN. On these points, both astronomical inputs produce coherent solar-like behavior, but the tidal-tensor forcing has the stronger physical interpretation.
	
The limitations of this result should also be stated explicitly. The model does not attempt to reproduce the detailed shape of each solar cycle, cycle asymmetries, stochastic emergence, meridional circulation, or the full spatial structure of the solar dynamo. Nor does the result prove a unique coupling pathway from planetary tides to the tachocline or convection zone. The experiment is narrower: it tests whether a weak, coherent, astronomically prescribed timing signal can modulate the phase and envelope of a self-excited nonlinear dynamo oscillator. In that restricted sense, Figure~\ref{fig:10} supports the weak-synchronization interpretation developed here, and the $2T_{xy}$ result makes this interpretation less dependent on an empirical ephemeris proxy and more directly connected to a physical weak-forcing quantity.

The main result of this comparison is therefore not that the reduced model provides a complete solar-dynamo mechanism, but that the synchronization signal initially identified with the empirical $RA_{\mathrm{tot}}(t)$ proxy persists when the forcing is replaced by the physically defined tidal-tensor descriptor $2T_{xy}(t)$. This makes the weak-synchronization interpretation less dependent on an ad hoc geometrical proxy and more directly connected to a mechanically defined planetary quantity.

\section{Discussion\label{sec:V}}
The main result of this study is the convergence of three independent levels of evidence. First, SSN, LOD, and planetary ephemeris descriptors share a common decadal-to-multidecadal pseudo-periodic structure, including 11 oscillatory components and a low-frequency trend. Second, this common structure is not limited to isolated spectral coincidences: the reconstructed modes also exhibit comparable envelope evolution and long-term phase organization. Third, a reduced nonlinear $\alpha$--$\Omega$ oscillator, when weakly forced by prescribed astronomical timing signals, reproduces the principal temporal organization of the historical sunspot record, including both Schwabe-scale pacing and broad multidecadal modulation. Importantly, this result is obtained not only with the empirical $RA_{\mathrm{tot}}$ phase proxy, but also with the physically defined quadrupolar tidal-tensor component $2T_{xy}$. Taken together, these results are consistent with, but do not prove, the interpretation that the solar cycle behaves as a self-excited nonlinear oscillator subject to weak external synchronization.

This interpretation is narrower than a claim of direct planetary driving. The results do not imply that planetary forcing powers the solar cycle or replaces the internal magnetic dynamo. Rather, they suggest that a coherent astronomical forcing may modulate the phase and long-term envelope of an oscillator whose energy source and primary instability remain internal to the Sun. In that sense, the present framework is fully compatible with standard dynamo theory: the solar cycle remains an internally generated magnetic oscillation, but not necessarily a strictly autonomous one.

The role of the barycentric description is more modest than a complete coupling mechanism. In a closed Sun--planet system, the barycentric motion of the Sun and the planetary orbital configuration are linked by angular-momentum conservation, and $L_{\odot}(t)$ is therefore a mechanically defined descriptor of the planetary angular-momentum organization. By contrast, $RA_{\mathrm{tot}}(t)$ is not a mechanical observable. It is used only as one empirical geometrical phase proxy among possible timing descriptors. The component $2T_{xy}(t)$ plays a different role: it is a physical component of the planetary tidal tensor and therefore provides a mechanically meaningful descriptor of the weak tidal timing imposed by the planetary configuration. The present interpretation therefore rests on weak synchronization by coherent astronomical timing, not on the claim that $RA_{\mathrm{tot}}(t)$ or the Sun's barycentric displacement directly forces the solar interior.

This distinction also clarifies why weak forcing is the relevant regime. A recurrent objection to planetary-solar coupling is that planetary tidal or gravitational effects are too small to compete directly with the forces governing the solar interior. That objection is decisive against any strong-driving hypothesis, but it is not decisive against phase or envelope modulation of a nonlinear oscillator. In many nonlinear systems, a forcing that is energetically negligible may still organize timing, synchronization windows, or long-term beating patterns. The present model should be read in that sense: not as a mechanism for supplying magnetic energy, but as a mechanism for modulating the temporal organization of an internally sustained cycle.

A magneto-Rossby relay provides one physically testable way to make this weak-forcing interpretation quantitative. Planetary tides at the tachocline radius remain extremely small, but if short-period spring-tide components overlap weakly damped tachocline Rossby or magneto-Rossby modes, the response can be amplified by an effective factor that includes damping, detuning, spatial projection, nonlinear saturation, and coupling to dynamo-relevant flows. The relevant requirement is not that the planets perform the work required to build the solar magnetic field; it is that a coherent tidal signal produces a sufficiently large modulation of a slow dynamo variable compared with its effective noise or phase-diffusion scale. This is the natural regime for stochastic resonance and noise-assisted synchronization (\eg \shortciteNP{Kramers1940,Gammaitoni1998,Hanggi1990,Budrikis2021,Albert2021}).

The reduced dynamo model supports this reading at the level of a minimal synchronization experiment. Its value does not lie in reproducing the absolute amplitude of sunspot number, nor in resolving the spatial complexity of the solar dynamo, but in showing that coherent astronomical timing can generate a solar-like temporal structure once passed through a nonlinear $\alpha$--$\Omega$ feedback loop. The model reproduces both the decadal recurrence of the Schwabe band and the broad envelope modulation associated with low- and high-activity intervals such as the Dalton Minimum and the Modern Maximum. This does not prove uniqueness of the proposed timing signal, but it shows that the weak-synchronization hypothesis is dynamically plausible within a nonlinear-oscillator framework. The comparable performance of $RA_{\mathrm{tot}}(t)$ and $2T_{xy}(t)$ is also informative: $RA_{\mathrm{tot}}$ appears to capture an empirical collective phase of planetary configuration, whereas $2T_{xy}$ expresses a corresponding physical quadrupolar tidal timing.

The LOD result is particularly important in this context. LOD is a terrestrial observable, not a solar one, and yet it shares the same ranked pseudo-periodic structure as SSN and the planetary quantities. This cross-domain agreement considerably strengthens the interpretation, because it reduces the likelihood that the observed structure is an artifact specific to one dataset or one extraction method. The significance of the result is therefore not only solar, but systemic: planetary, solar, and terrestrial records appear to share a common low-frequency timing architecture.

Several objections nevertheless remain legitimate and should be considered explicitly. The first concerns the apparent absence of simple direct correlations between solar activity indicators and planetary configurations, since some studies have reported negative or inconclusive results (\eg \shortciteNP{Seker2013,Nataf2022,Weisshaar2023,Cionco2023}). Such findings argue against any naive one-to-one geometric triggering or direct tidal-driving scenario, but they do not exclude a weaker form of synchronization acting on the phase or envelope of a nonlinear solar oscillator. A second objection concerns the possibility of statistical coincidence. Here, the argument does not rest on one or two isolated matching periods, but on the joint recurrence of a ranked set of decadal-to-multidecadal pseudo-periods, on their temporal reconstruction, and on the agreement of their cumulative distributions. Nevertheless, the statistical tests presented here should be regarded as diagnostic rather than definitive; future work should include red-noise surrogates, phase-randomized surrogates, and full-pipeline SSA tests. A third objection concerns the use of SSA. In the present work, SSA is employed only as a decomposition tool, not as a generative model. A final objection concerns the simplicity of the reduced dynamo model. This criticism is justified insofar as the model omits spatial transport, stochastic emergence, meridional circulation, and many details of realistic solar-dynamo physics. But this simplicity is also intentional: the model is not meant to serve as a full solar simulator, only as a minimal dynamical test of whether weak synchronization by astronomical timing is plausible. The tidal-tensor comparison addresses a further objection: the main dynamical result is not dependent solely on an arbitrary geometrical sum of right ascensions.

The present framework also makes falsifiable predictions. If planetary synchronization is relevant, the phase relationship between the solar cycle and the planetary forcing should remain slowly varying and bounded rather than becoming erratic. Future solar cycles should preserve, at least statistically, the decadal and multidecadal organization encoded in the forcing signal. In addition, the common pseudo-periodic modes identified in SSN, LOD, and planetary observables should remain structurally robust under reasonable variations of decomposition parameters. Conversely, a systematic loss of coherence, disappearance of the ranked common bands, or failure of the forced dynamo to maintain the observed phase-envelope organization would count against the present interpretation.

The main limitation of this study is therefore quantitative and physical. The dynamical framework is plausible, the empirical agreement is suggestive, and the reduced model is successful at the level of temporal organization; however, the precise physical pathway by which an ephemeris-based timing signal could couple to solar internal shear remains incompletely specified. The magneto-Rossby relay should therefore be viewed as a quantitative hypothesis rather than as a demonstrated mechanism. It requires that the short-period tidal spectrum couple to tachocline Rossby or magneto-Rossby modes with sufficient effective gain, and that the resulting modulation influence a dynamo-relevant coefficient or flow in the presence of convective and magnetic noise. This point should now be addressed by future work using more realistic dynamo models, helioseismic constraints on tachocline variability, and prospective tests on future solar cycles. The purpose of the present paper is not to close that question, but to show that it can be posed in a physically cautious, empirically motivated, and falsifiable form.

We therefore conclude that the most cautious reading of the combined results is not that planetary forcing replaces the solar dynamo, but that coherent astronomical timing may weakly synchronize or phase-modulate it. Within this interpretation, the solar cycle remains an internally sustained nonlinear magnetic oscillator whose phase and long-term amplitude modulation may be partially organized by external timing. Such a view connects the observational periodicities, the barycentric angular-momentum descriptor, the physical tidal-tensor descriptor, and the reduced forced-oscillator results within a single nonlinear dynamical hypothesis, while leaving the detailed coupling mechanism open.

\section{Conclusion\label{sec:VI}}
In this study, we have examined whether the solar cycle can be interpreted as a nonlinear self-excited oscillator susceptible to weak astronomical synchronization. Three main results motivate this view. First, sunspot number, terrestrial length-of-day, and planetary ephemeris descriptors share a common decadal-to-multidecadal pseudo-periodic structure. Second, this agreement extends beyond isolated spectral peaks to the temporal organization of the reconstructed modes themselves. Third, a reduced nonlinear $\alpha$--$\Omega$ oscillator, weakly forced by prescribed timing signals constructed from the empirical phase proxy $RA_{\mathrm{tot}}(t)$ and from the physical tidal-tensor descriptor $2T_{xy}(t)$, reproduces the principal Schwabe-scale pacing and the broad multidecadal modulation of the historical sunspot record. The $2T_{xy}$ result is central because it shows that the synchronization is recovered with a physically defined quadrupolar tidal quantity rather than only with a geometrical right-ascension proxy.

These results do not imply that planetary forcing replaces the internal solar dynamo or provides the energy source of the solar cycle. Rather, they support a more specific and cautious interpretation: the solar dynamo behaves as an internally sustained nonlinear oscillator whose phase and long-term envelope may be modulated by weak, coherent astronomical timing. In that sense, the present framework remains compatible with standard dynamo physics while relaxing the assumption that the solar cycle is strictly autonomous.

The broader significance of this work lies in connecting planetary, solar, and terrestrial variability within a nonlinear-dynamical picture based on common timing structure, a mechanically defined barycentric angular-momentum descriptor, a physically defined tidal-tensor descriptor, and the response of a reduced oscillator to weak forcing. The framework also yields falsifiable expectations: persistence of bounded phase coherence, robustness of the common pseudo-periodic modes, and continued reproducibility of large-scale envelope modulation in future solar cycles. Whether future observations strengthen or weaken this interpretation, they will directly clarify the extent to which celestial mechanics contributes, if at all, to long-term solar variability.

Future work should now move toward more realistic dynamical formulations, including higher-dimensional dynamo models, helioseismic constraints on rotational shear, explicit tests of tachocline Rossby and magneto-Rossby response to planetary tidal spectra, and prospective tests on future solar cycles. In particular, the tidal-tensor descriptor $2T_{xy}(t)$ should be tested together with spring-tide forcing, mode damping, nonlinear beating, and stochastic phase diffusion in models closer to tachocline dynamics. The purpose of the present study is therefore not to close the question of planetary-solar coupling, but to show that it can be posed in a physically cautious and empirically testable way within the framework of nonlinear synchronization.

\appendix

\section{Statistical support for the ranked common periodicities\label{app:A}}
In this appendix, we quantify the degree of agreement between the ranked pseudo-periodicities identified in SSN, LOD, and $L_{\odot}$, and compare them with the reference set $LR^{*}$ associated with the planetary Laplace-type resonance timescales listed in the last column of Table~\ref{tab:02}. The purpose is to test whether the observed similarity between these ranked period sets can plausibly arise by chance.

\subsection{Pairwise correlations and bootstrap test}

The ranked periodicities extracted from SSN, LOD, $L_{\odot}$, and $LR^{*}$ are strongly correlated with one another. Table~\ref{tab:C01} reports the corresponding pairwise correlation coefficients. In particular, the correlations among SSN, LOD, and $L_{\odot}$ are all very close to unity, while the correlations with the reference set $LR^{*}$ also remain high.

To evaluate whether such agreement could emerge accidentally from ordered random series, we generated 3400 bootstrap realizations under a null hypothesis in which each set consists of eleven exponentially distributed random periods. Each synthetic set was ranked in increasing order and then compared pairwise using the same correlation statistic as for the observed data. The resulting empirical probabilities of obtaining correlations equal to or greater than the observed ones are listed above the diagonal in Table~\ref{tab:C01}.

\begin{table}[H]
	\centering
	\caption{Pairwise statistics for the ranked periodicities. Correlation coefficients are given below the diagonal. The diagonal gives the number of ranked periods in each series. Entries above the diagonal are empirical probabilities of obtaining an equal or larger correlation coefficient under the bootstrap null hypothesis of ordered exponentially distributed random series.}
	\label{tab:C01}
	
	\vspace{0.5em}
	
	{
		\renewcommand{\arraystretch}{1.22}
		\setlength{\tabcolsep}{10pt}
		
		\begin{tabular}{lcccc}
			\toprule
			& \textbf{SSN}
			& \textbf{LOD}
			& \textbf{$L_{\odot}$}
			& \textbf{$LR^{*}$} \\
			\midrule
			
			\textbf{SSN}
			& \cellcolor{gray!20} 11
			& 0.000294
			& 0.000294
			& 0.000588 \\
			
			\textbf{LOD}
			& 0.999154
			& \cellcolor{gray!20} 11
			& 0.000294
			& 0.001765 \\
			
			\textbf{$L_{\odot}$}
			& 0.998874
			& 0.998874
			& \cellcolor{gray!20} 11
			& 0.004706 \\
			
			\textbf{$LR^{*}$}
			& 0.995975
			& 0.995629
			& 0.993430
			& \cellcolor{gray!20} 11 \\
			
			\bottomrule
		\end{tabular}
	}
	
\end{table}

The observed correlations among SSN, LOD, and $L_{\odot}$ exceed almost all values obtained under the null model. The correlations involving $LR^{*}$ are slightly less extreme, but still remain highly unlikely under the same null hypothesis. These results indicate that the ranked periodic agreement is not easily explained by accidental matching of independently drawn ordered period sets.

\subsection{Comparison of empirical cumulative distributions}

A complementary way to assess the similarity of the ranked period sets is to compare their empirical cumulative distribution functions. Figure~\ref{fig:06} in the main text shows that the cumulative distributions of SSN, LOD, $L_{\odot}$, and $LR^{*}$ are closely aligned over the full ranked range.

To quantify this agreement, we applied the two-sample Kolmogorov--Smirnov (KS) test pairwise to the ranked periods extracted from SSN, LOD, $L_{\odot}$, and $LR^{*}$. In all cases, the maximum absolute difference between empirical cumulative distributions remains small. For the observed ranked sets of size 11, the KS distance satisfies
\begin{equation}
	D_{11,11}<0.1,
	\label{eq:C01}
\end{equation}
which implies
\begin{equation}
	\lambda_{\mathrm{KS}} < 0.1 \left(\frac{11}{2}\right)^{1/2} \approx 0.2345.
	\label{eq:C02}
\end{equation}
This corresponds to a large KS p-value, so the null hypothesis that the compared ranked samples originate from the same continuous distribution cannot be rejected. In other words, the KS test does not support the interpretation that these ranked periodicities arise from unrelated parent distributions.

\subsection{Interpretation}

Taken together, these diagnostics show that the ranked pseudo-periodicities identified in SSN, LOD, and $L_\odot$ are unusually similar under the simple null models considered here. This does not prove a common physical origin. It indicates only that the observed agreement is not well represented by independent ordered random period sets, and motivates more stringent surrogate tests in future work.

This statistical result does not, by itself, identify the physical mechanism responsible for the common structure. However, it provides an important constraint on interpretation. If the ranked periodicities in the solar, terrestrial, and planetary observables were generated by independent processes, their level of agreement would be difficult to explain under the null models considered here. The most natural reading is therefore that these observables share a common organizing constraint, which in the present work is interpreted as being consistent with weak synchronization by coherent astronomical timing.

\section{Reduced-dynamo parameters and robustness\label{app:B}}

{The reduced dynamo used in Section~4 is intended as a minimal synchronization experiment rather than as a calibrated model of the solar interior. For completeness, we report here the numerical coefficients used in the integrations shown in Figure~\ref{fig:10}, the intrinsic damped natural period of the unforced reduced oscillator, and a local robustness check of the $2T_{xy}$-forced solution.

\begin{table}[htbp]
	\centering
	\caption{Numerical coefficients and intrinsic timescale diagnostics of the reduced dynamo. The coefficient $\alpha_0$ is reported as the magnitude of the effective $\alpha$--$\Omega$ feedback coefficient; the intrinsic period is evaluated using the oscillatory sign convention of the reduced feedback loop.}
	\label{tab:appendix-dynamo-parameters}
	\begin{tabular}{lll}
		\hline
		Quantity & Value & Meaning \\
		\hline
		$\Omega_0$ & $0.60$ & baseline $\Omega$-effect coefficient \\
		$\alpha_0$ & $0.54$ & baseline $\alpha$-effect coefficient \\
		$\eta_A$ & $0.12$ & effective damping of the poloidal amplitude \\
		$\eta_B$ & $0.10$ & effective damping of the toroidal amplitude \\
		$\mu_A$ & $0.08$ & nonlinear saturation of the poloidal amplitude \\
		$\mu_B$ & $0.20$ & nonlinear saturation of the toroidal amplitude \\
		\hline
		$P_0$ & $11.040$ yr & intrinsic damped natural period \\
		$2/(\eta_A+\eta_B)$ & $9.091$ yr & linear damping time \\
		$P_{2T_{xy}}$ & $11.126$ yr & dominant period of the $2T_{xy}$-forced solution \\
		\hline
	\end{tabular}
\end{table}

For the adopted oscillatory $\alpha$--$\Omega$ convention, the linearized unforced system has the natural angular frequency,
\[
\omega_0 =
\left[
|\Omega_0\alpha_0|
-
\left(\frac{\eta_B-\eta_A}{2}\right)^2
\right]^{1/2},
\]

so that $P_0=2\pi/\omega_0=11.040$ yr. Direct autonomous integrations from several initial conditions recover this early damped oscillation with the same period. Since the oscillation decays in the absence of prescribed timing, the reduced model is used here as a weakly forced synchronization experiment, not as a self-sustained autonomous solar-cycle model.

\begin{table}[htbp]
	\centering
	\small
	\caption{One-at-a-time robustness check of the $2T_{xy}$-forced reduced dynamo. Each coefficient was varied separately by $\pm 20\%$, while the input normalization and comparison procedure were kept unchanged. The reported diagnostics are computed from the 2 yr smoothed toroidal output, as in Figure~\ref{fig:10}.}
	\label{tab:appendix-dynamo-robustness}
	\begin{tabular}{lccccc}
		\hline
		Case & Scale & $r$ & $\mathrm{RMSE}_{z}$ & $\mathrm{MAE}_{z}$ & Period (yr) \\
		\hline
		Reference & $1.00$ & $0.660418$ & $0.824031$ & $0.682133$ & $11.126$ \\
		$\Omega_0+20\%$ & $1.20$ & $0.660425$ & $0.824023$ & $0.682124$ & $11.126$ \\
		$\Omega_0-20\%$ & $0.80$ & $0.660412$ & $0.824039$ & $0.682141$ & $11.126$ \\
		$\alpha_0+20\%$ & $1.20$ & $0.660421$ & $0.824027$ & $0.682130$ & $11.126$ \\
		$\alpha_0-20\%$ & $0.80$ & $0.660415$ & $0.824034$ & $0.682136$ & $11.126$ \\
		$\eta_A+20\%$ & $1.20$ & $0.660418$ & $0.824031$ & $0.682133$ & $11.126$ \\
		$\eta_A-20\%$ & $0.80$ & $0.660418$ & $0.824030$ & $0.682132$ & $11.126$ \\
		$\eta_B+20\%$ & $1.20$ & $0.660415$ & $0.824034$ & $0.682136$ & $11.126$ \\
		$\eta_B-20\%$ & $0.80$ & $0.660421$ & $0.824027$ & $0.682129$ & $11.126$ \\
		$\mu_A+20\%$ & $1.20$ & $0.660417$ & $0.824032$ & $0.682135$ & $11.126$ \\
		$\mu_A-20\%$ & $0.80$ & $0.660420$ & $0.824028$ & $0.682130$ & $11.126$ \\
		$\mu_B+20\%$ & $1.20$ & $0.660420$ & $0.824029$ & $0.682131$ & $11.126$ \\
		$\mu_B-20\%$ & $0.80$ & $0.660415$ & $0.824035$ & $0.682134$ & $11.126$ \\
		\hline
	\end{tabular}
\end{table}

Across all twelve one-at-a-time perturbations, the smoothed correlation remains between $0.660412$ and $0.660425$, $\mathrm{RMSE}_{z}$ remains between $0.824023$ and $0.824039$, $\mathrm{MAE}_{z}$ remains between $0.682124$ and $0.682141$, and the dominant period remains close to $11.126$ yr. This local robustness check is not a full exploration of parameter space, but it shows that the comparison in Figure~\ref{fig:10} is not tied to fine tuning of a single reduced-dynamo coefficient.
	
The near invariance of $r$, $\mathrm{RMSE}_{z}$, and $\mathrm{MAE}_{z}$ under these perturbations should be interpreted as robustness of the normalized timing and envelope diagnostics, not as invariance of the dimensional dynamo amplitude. A separate stress test confirms this interpretation. In that test, $\Omega_0$ and $\alpha_0$ were doubled, $\eta_A$ and $\eta_B$ were halved, and $\mu_A$ and $\mu_B$ were doubled. The raw and smoothed $B(t)$ responses changed by a relative norm of $0.206$, whereas the corresponding $z$-scored smoothed waveform changed by only $2.54\times10^{-4}$, with a maximum absolute difference of $0.00423$. Thus, the reduced coefficients affect primarily the amplitude scale of the response, while the normalized phase and envelope agreement are mainly controlled by the prescribed $2T_{xy}(t)$ timing input.

\nocite{Morth1979}
\bibliographystyle{fchicago}
\bibliography{ssn_dynamo_biblio}

\end{document}